\documentclass[sigconf]{acmart}
\setcopyright{rightsretained}
\copyrightyear{2020}
\acmYear{2020}
\acmConference[CASCON'20]{CASCON'20, November 10 - 13 2020}{November 10 - 13 2020}{Toronto, Canada}
\acmBooktitle{}
\acmPrice{}
\acmISBN{}
\acmDOI{}

\AtBeginDocument{%
  \providecommand\BibTeX{{%
    \normalfont B\kern-0.5em{\scshape i\kern-0.25em b}\kern-0.8em\TeX}}}
    
\usepackage{pifont}
\newcommand{\cmark}{\ding{51}}%
\newcommand{\xmark}{\ding{55}}%
\usepackage{subfig}
\usepackage{lscape}
\usepackage{listings, xcolor}
\definecolor{verylightgray}{rgb}{.97,.97,.97}
\usepackage{color}
\definecolor{lightgray}{rgb}{0.95, 0.95, 0.95}
\definecolor{darkgray}{rgb}{0.4, 0.4, 0.4}
\definecolor{editorGray}{rgb}{0.95, 0.95, 0.95}
\definecolor{editorOcher}{rgb}{1, 0.5, 0} 
\definecolor{editorGreen}{rgb}{0, 0.5, 0} 
\definecolor{orange}{rgb}{1,0.45,0.13}		
\definecolor{olive}{rgb}{0.17,0.59,0.20}
\definecolor{brown}{rgb}{0.69,0.31,0.31}
\definecolor{purple}{rgb}{0.38,0.18,0.81}
\definecolor{lightblue}{rgb}{0.1,0.57,0.7}
\definecolor{lightred}{rgb}{1,0.4,0.5}
\usepackage{upquote}
\usepackage{listings}
\lstdefinelanguage{Solidity}{
	keywords=[1]{anonymous, assembly, assert, balance, break, call, callcode, case, catch, class, constant, continue, constructor, contract, debugger, default, delegatecall, delete, do, else, emit, event, experimental, export, external, false, finally, for, function, gas, if, implements, import, in, indexed, instanceof, interface, internal, is, length, library, log0, log1, log2, log3, log4, memory, modifier, new, payable, pragma, private, protected, public, pure, push, require, return, returns, revert, selfdestruct, send, solidity, storage, struct, suicide, super, switch, then, this, throw, transfer, true, try, typeof, using, value, view, while, with, addmod, ecrecover, keccak256, mulmod, ripemd160, sha256, sha3}, 
	keywordstyle=[1]\color{blue}\bfseries,
	keywords=[2]{address, bool, byte, bytes, bytes1, bytes2, bytes3, bytes4, bytes5, bytes6, bytes7, bytes8, bytes9, bytes10, bytes11, bytes12, bytes13, bytes14, bytes15, bytes16, bytes17, bytes18, bytes19, bytes20, bytes21, bytes22, bytes23, bytes24, bytes25, bytes26, bytes27, bytes28, bytes29, bytes30, bytes31, bytes32, enum, int, int8, int16, int24, int32, int40, int48, int56, int64, int72, int80, int88, int96, int104, int112, int120, int128, int136, int144, int152, int160, int168, int176, int184, int192, int200, int208, int216, int224, int232, int240, int248, int256, mapping, string, uint, uint8, uint16, uint24, uint32, uint40, uint48, uint56, uint64, uint72, uint80, uint88, uint96, uint104, uint112, uint120, uint128, uint136, uint144, uint152, uint160, uint168, uint176, uint184, uint192, uint200, uint208, uint216, uint224, uint232, uint240, uint248, uint256, var, void, ether, finney, szabo, wei, days, hours, minutes, seconds, weeks, years},	
	keywordstyle=[2]\color{teal}\bfseries,
	keywords=[3]{block, blockhash, coinbase, difficulty, gaslimit, number, timestamp, msg, data, gas, sender, sig, value, now, tx, gasprice, origin},	
	keywordstyle=[3]\color{violet}\bfseries,
	identifierstyle=\color{black},
	sensitive=false,
	comment=[l]{//},
	morecomment=[s]{/*}{*/},
	commentstyle=\color{gray}\ttfamily,
	stringstyle=\color{red}\ttfamily,
	morestring=[b]',
	morestring=[b]"
}
\lstdefinelanguage{JavaScript}{
  morekeywords={typeof, new, true, false, catch, function, return, null, catch, switch, var, if, in, while, do, else, case, break},
  morecomment=[s]{/*}{*/},
  morecomment=[l]//,
  morestring=[b]",
  morestring=[b]'
}
\usepackage[T1]{fontenc}
\usepackage{listings,xcolor,beramono}
\usepackage{tikz}
\UseRawInputEncoding
\makeatletter
\newenvironment{btHighlight}[1][]
{\begingroup\tikzset{bt@Highlight@par/.style={#1}}\begin{lrbox}{\@tempboxa}}
{\end{lrbox}\bt@HL@box[bt@Highlight@par]{\@tempboxa}\endgroup}

\newcommand\btHL[1][]{%
  \begin{btHighlight}[#1]\bgroup\aftergroup\bt@HL@endenv%
}
\def\bt@HL@endenv{%
  \end{btHighlight}%
  \egroup
}
\newcommand{\bt@HL@box}[2][]{%
  \tikz[#1]{%
    \pgfpathrectangle{\pgfpoint{1pt}{0pt}}{\pgfpoint{\wd #2}{\ht #2}}%
    \pgfusepath{use as bounding box}%
    \node[anchor=base west, fill=orange!30,outer sep=0pt,inner xsep=1pt, inner ysep=0pt, rounded corners=3pt, minimum height=\ht\strutbox+1pt,#1]{\raisebox{1pt}{\strut}\strut\usebox{#2}};
  }%
}
\makeatother

\lstdefinestyle{Solidity}{
    language={Solidity}, 
     moredelim=**[is][\btHL]{`}{`},
    moredelim=**[is][{\btHL[fill=green!30]}]{@}{@},
    moredelim=**[is][{\btHL[fill=cyan!30]}]{~}{~},
    extendedchars=true,
        }
\usepackage{qtree}
\usepackage{multicol}        
\def\BibTeX{{\rm B\kern-.05em{\sc i\kern-.025em b}\kern-.08em
    T\kern-.1667em\lower.7ex\hbox{E}\kern-.125emX}}

\begin{document}

\title{A Survey of Security Vulnerabilities in Ethereum Smart
Contracts}

\author{Noama Fatima Samreen,   Manar H. Alalfi}
\email{noama.samreen,   manar.alalfi@ryerson.ca}
\affiliation{%
  \institution{Department of Computer Science, Ryerson University}
  \city{Toronto}
  \state{ON, Canada}
  }


\begin{abstract}
Ethereum Smart Contracts based on Blockchain Technology (BT) enables monetary transactions among peers on a blockchain network independent of a central authorizing agency. Ethereum Smart Contracts are programs that are deployed as decentralized applications, having the building blocks of the blockchain consensus protocol. This enables consumers to make agreements in a transparent and conflict-free environment. However, there exists some security vulnerabilities within these smart contracts that are a potential threat to the applications and their consumers and have shown in the past to cause huge financial losses. In this study, we review the existing literature and broadly classify the BT applications. As Ethereum smart contracts find their application mostly in e-commerce applications, we believe these are more commonly vulnerable to attacks. In these smart contracts, we mainly focus on identifying vulnerabilities that programmers and users of smart contracts must avoid. This paper aims at explaining eight vulnerabilities that are specific to the application level of BT by analyzing the past exploitation case scenarios of these security vulnerabilities. We also review some of the available tools and applications that detect these vulnerabilities in terms of their approach and effectiveness. 
We also investigated the availability of detection tools for identifying these security vulnerabilities and lack thereof to identify some of them.
\end{abstract}




\maketitle

\section{Introduction}
Attributing to the wide range applicability of Blockchain Techhnology(BT), it has been finding popularity in many domains. Bitcoin was the first version of cryptocurrency applied using BT \cite{Bitcoin} and has since been used in many other applications such as e-commerce, trade and commerce, production and manufacturing, banking, and gaming. BT uses a peer-to-peer (peers are known as miners in BT) framework which is a more decentralized approach to storing transactions and data registers. As there is no single point of failure or a third-party centralized control of transactions, BT has been standing out from cryptocurrency-based other technologies. It uses a chain of blocks in which each block is locked cryptographically using the hash of the previous block it is linked to, which creates an immutable database of all transactions stored as a digital ledger, and it cannot be changed without affecting all the blocks linked together in the chain \cite{surveyattacks}. However, recent research to identify the existence of security vulnerabilities in Ethereum Smart Contracts have shown that many applications have been exposed to attacks because of vulnerabilities found in application level of Ethereum Smart Contracts \cite{vulnerabilities}. 

Ethereum Smart Contracts are typically written in Solidity Language and this paper presents a classification for vulnerabilities in these Solidity-based Ethereum Smart Contracts. Our methodology can be characterized as targeting security from three perspectives: vulnerabilities, exploitation case studies, and preventive techniques. 
For each vulnerability, we discuss, among other things, its research statistics (i.e., detection tools available to identify the vulnerability, analysis method most preferred by researchers to identify the vulnerability). For each exploitation case, we discuss, among other things, the vulnerability exploited, tactic, and financial losses incurred in terms of ether. For each preventive technique, we discuss its mechanism and the vulnerability it aims to protect from exploitation. 
We aim to provide future research directions by providing statistics of research done on each vulnerability to address the severity of a vulnerability and the requirement of further work on open problems.
    
We identify eight application level security vulnerabilities and classify them according to the NIST’s Bugs Framework \cite{NIST}. To do so, we collected information from the NIST’s website to match the descriptions of existing bug classes with the Ethereum Smart Contracts security vulnerabilities discussed. As highlighted in Table[\ref{vulDet}], most of these vulnerabilities are classified as \textit{Not Available (NA)} concerning NIST-BF classification. This is because most of these vulnerabilities are specific to developmental methodology of a Solidity based Ethereum Smart Contract and BT and do not match an exact \textit{Bugs Framework (BF) category} outlined by NIST. This classification concerning NIST categorization of software bugs will aim at a better understanding of the nature of each of the Ethereum Smart Contract's application level security vulnerabilities. 

\section{Background}
\subsection{Blockchain Technology (BT)}
Based on the industry they target, BT applications can be broadly classified into three categories: Public Blockchain 1.0, Public Blockchain 2.0 and Private Blockchain 3.0. 
 \begin{itemize}
     \item Public Blockchain 1.0 is used predominantly in the finance industry for digital payments and currency transfers.
     \item Public Blockchain 2.0 is used in the e-commerce industry and it includes Ethereum Smart Contracts, such applications process financial contracts intelligently and provide a foundation for digital asset ownership. As Ethereum smart contracts find their application mostly in e-commerce applications, we believe these are more commonly vulnerable to attacks caused by simple coding errors. 
     
     In this paper, we mainly focus on identifying vulnerabilities in these smart contracts that programmers and users of smart contracts must avoid. Case studies outline that Ethereum smart contracts are vulnerable to simple coding errors such as re-entrance (recursive calling of functions: A calling B while B is calling A), wrong constructor name, typecasts, unintended function exposure, stack overflow, etc. as this generation of BT is public and distributed. These coding errors can aid an attacker in manipulating transactions to successfully launch an intrusion attack by techniques such as publishing malicious contract on the BT network to receive more transactions than it sends out thereby collecting Ether (digital asset bearer or a token by which applications are processed on an Ethereum network) many times over within a single transaction, exploiting the visibility modifiers to misuse function delegation etc. 
 \item Private Blockchain 3.0 is used in the government, health, science industries and are hence considered private.
 \end{itemize}
\subsection{NIST Bugs Framework}
In this paper, we leverage the National Institute of Standards and Technologies Bugs (NIST’S) Framework to provide another classification of these vulnerabilities to provide a basis of comparison with other common software bugs reported by NIST\cite{NIST}. This categorization by NIST is an effort to accurately describe a commonly occurring software bug and it incorporates definitions and attributes of each software bug class along with their causes and consequences. This classification provides researchers and developers to match vulnerabilities in new technologies to the previously researched bugs and adopt appropriate preventive techniques.
\subsection{Analysis Methodologies}
Literature review of the selected Smart Contracts vulnerabilities detection tools/frameworks for this paper shows that these tools/frameworks adopted one of the following methods of smart contract analysis, 
\subsubsection{Static Analysis}
\begin{enumerate}
    \item Symbolic Execution - The execution of code using symbols rather than real values for the variables. This analysis results in algebraic terms of operating these symbols and  the conditional statements in the program result in propositional formulas that direct the flow of execution. The feasibility of a path of flow is determining if the conjunction of formulas on the path is satisfied.
    \item CFG (Control Flow Graph) Construction - The representation of a program in a directed graph. An edge of this graph represents the flow of execution with conditions mentioned on the edge as a label.
    \item Pattern Recognition - The classification of a program's basics units or data depending on the prior knowledge or statistical information gained from patterns.
    \item Rule-based Analysis - The checking/analyzing of code against a rule-based specification of its behaviour. The rule-based specification describes scenarios during execution and enforces constraints on the sequence of operations and data inputs.
    \item De-compilation Analysis - The representation of Ethereum Virtual Machine (EVM) bytecode  with a higher abstraction level to improve the parsing of the code and data flow analysis.
\end{enumerate}
\subsubsection{Dynamic Analysis}
\begin{enumerate}
    \item Execution Trace at Run time - Tracing the sequence of instructions that are executed during a particular run of the code
    \item Fuzzing Input Generation - Fuzzing is an automated analysis method that tests program execution by providing structured data as inputs to a computer program. The program under analysis is then monitored for unexpected behaviour such as unusual code path, or crashes.
\end{enumerate}
\section{Related Work}
We conducted a literature review and we inferred that there is a lack of a comprehensive study of these security vulnerabilities in Solidity based Ethereum Smart Contracts.
A survey of vulnerabilities by L. Luu et al. \cite{making} focuses only on the security vulnerabilities that exist in the Ethereum Smart Contracts without providing a detailed study of their exploitation cases or preventive techniques. Another study by N. Atzei et al. \cite{sok} discusses security vulnerabilities and some of their real world attack scenarios in general without providing a mapping between the vulnerabilities and the attacks. 
Then there is a survey of the current research available in the field of ethereum smart contract by Alharby and Moorsel. \cite{mappingStudy} that characterizes the surveys available depending on their nature of survey as the surveys identifying codifying issues, security issues, privacy issues, and performance issues.
One of the most recent surveys in this area is by Praitheeshan et al. \cite{surveyLatest} which lists 13 security vulnerabilities (involving application level, BT level and Ethereum Virtual Machine-EVM level security vulnerabilities) in Ethereum Smart Contracts but maps them to only four major attacks scenarios. Therefore, it fails to give an example pattern or a real world exploitation case scenario for each of the security vulnerability discussed.

On the other hand, there has been extensive development of automated tools in the industry to detect these vulnerabilities. These developers utilize the research available in this area to produce highly efficient state-of-the-art tools to detect these vulnerabilities. 
Some of the research work in developing automated tools for detecting these vulnerabilities focuses on detecting only a speciﬁc type of vulnerability without analyzing the vulnerability in detail with its exploitation cases and preventive techniques. \cite{Reguard},\cite{Oyente}, \cite{ContractFuzzer} 
One such recent tool is ETHPLOIT by Q. Zhang et al. \cite{Ethploit} to automatically detect vulnerabilities that have been exploited in Ethereum smart contracts. This tool adopts light-weight techniques to answer the problems of previously developed tools. These problems consisted of unsolvable constraints and Blockchain effects. It is claimed by Q. Zhang et al. \cite{Ethploit} that this tool achieves precise and efficient smart contract analysis and successfully detects more exploits than previous exploit generation tools.

However, a new research paper SMARTSHIELD by Y. Zhang et al. \cite{smartshield} utilizes EVM bytecode analysis and provides an automatic correction mechanism to avoid vulnerable patterns in Ethereum smart contracts. It does this rectification by extracting EVM bytecode level semantic data to transform the vulnerable smart contracts into secure ones. 
And then, there are also some surveys like the one by Angelo and Salzer \cite{SurveyTools} that compare only the detection tools available in the market without actually discussing the characteristics of the vulnerabilities these tools excel or fail at detecting.

One of the most recent surveys published in Feb. 2020 by Durieux et al. covers three research questions regarding these automated tools available in the market. This research work by Durieux et al. questions the effectiveness of these tools in terms of precision in detecting these vulnerabilities in Ethereum Smart Contracts. Durieux et al. next articulates about the quantitative analysis of the vulnerabilities present in the Ethereum blockchain main network. 

Different from the existing surveys discussed above, our paper aims to particularly analyse the vulnerability pattern, their real world exploitation cases, their preventive techniques and the detection methods adopted by currently available tools for analyzing ethereum smart contracts. We highlight the need for a comprehensive study on the security analysis methods of vulnerable smart contracts on the Ethereum platform. 
Furthermore, this paper is different from the existing ethereum smart contracts surveys because we investigate each security vulnerability in detail along with the available detection tools to analyse the security vulnerability and the methodology adopted by these detection tools to identify this vulnerability in ethereum smart contracts.

\section{Vulnerabilities}
This paper combines the identification and analysis of eight of the application level security vulnerabilities along with their real-world exploitation cases to better capture the vulnerabilities scenarios in Solidity-based Ethereum Smart Contracts. Table[\ref{vulDet}] shows the vulnerable contracts used in exploitation cases, their preventive techniques respectively along with the matching NIST bug class for each vulnerability.
\begin{table*}
\caption{Smart contract vulnerabilities, their preventive techniques and their NIST bug classification;NIST-BF - National Institute of Standards and Technology- Bugs Framework; NA - Not Available, UCE - Unchecked Error, ARC - Arithmetic or Conversion Fault}
\label{vulDet}
 \scalebox{0.95}{
 \begin{tabular}{|p{0.5cm}|p{3cm}|p{4cm}|p{3cm}|p{3cm}|p{4cm}|} 
 \hline
 S.no. & Contract Name & Vulnerability & Vulnerability Level & NIST-BF Class & Preventive Technique \\ 
 \hline
 1. & The DAO & Re‐entrancy(recursive‐calling vulnerability: A calling B calling A)& Security & NA &
 Placing external call logic as the last piece of code in a program\\
\hline
2. & King of the ether & Out-of-Gas Exception Handling & Functional & NA & use \textit{transfer()} instead of \textit{send()} \\
\hline
3. & Governmental (Ponzi Scheme) & Unpredictable state due to mishandled exceptions & Security/Functional & NA &  Updating Solidity Language to handle exceptions in a uniform manner is required \\
\hline
4. & Second Parity MultiSig Wallet & Call-to-Unknown vulnerability & Security & UCE & Making stateless libraries of vulnerable contracts to avoid external state changing of the contract\\
\hline
5. & Reentrancy Honey Pot & Typecast vulnerability & Developmental & NA & Using \textit{new} to create an instance of referenced contract\\
\hline
6. & Odd and Even Game & Weak Field Modifiers vulnerability & Developmental & NA & Using \textit{internal} to protect information leakage\\
\hline
7. & Proof of Weak Hands Coin (PoWHC) & Integer Underflow/Overflow vulnerability & Arithmetic or Conversion Fault & ARC & Using mathematical libraries instead of the standard math operations (addition, subtraction and multiplication)\\
\hline
8. & HYIP & DoS by external call vulnerability & Unchecked Error Class & UCE & Asking recipient to \textit{pull} funds out rather than sender using \textit{push} to send out the funds.
 
 Removing dependence of conditional statements or iterational statements on an external call.\\
\hline
\end{tabular}}
\end{table*}
\subsection{Reentrancy}
 A reentrancy attack can drain a smart contract of its ether, can aid an intrusion into the contract code. When an external call function to another untrustworthy contract is made and an attacker gains control of this untrustworthy contract, they can make a recursive call back to the original function, unexpectedly repeating transactions that would have otherwise not run, and eventually consume all the gas.
 \subsubsection*{Exploitation Case of Reentrancy Vulnerability - The DAO Attack}
 The Decentralized Autonomous Organization (known as the DAO) was initiated in May 2016 as a venture capital fund for the crypto and decentralized space\cite{surveyattacks}. During the creation period of the DAO, anyone could send Ether to a unique wallet address in exchange for DAO tokens. 
 Anyone with DAO tokens could vote on the pitch and receive rewards in return if the projects turned a profit. However, on June 17, 2016, a hacker was able to attack this Smart contract by exploiting a vulnerability in the code that allowed him to transfer funds from the DAO. As reported by M. Saad et al.\cite{DBLP:journals/corr/abs-1904-03487} approximately, 3.6 million Ether was stolen, the equivalent of USD 70M at the time. 
The reentrancy vulnerability exploitation in the DAO attack(as shown in Listing \ref{daoattackLst}) was accomplished in four steps,
\begin{itemize}
    \item The Attacker initiates a transaction by calling withdraw function of Victim;
    \item The Victim transfers the money and calls the fallback function of the Attacker;
    \item The fallback function recursively calls the withdraw function again, i.e., \textit{Reentrancy};
    \item Within an iteration bound, extra ether will be transferred multiple times to the Attacker.
\end{itemize}

\begin{lstlisting}[style=Solidity, basicstyle=\fontsize{7}{8}\selectfont, caption={Simplified DAO Attack - Reentrancy Vulnerability} ,label={daoattackLst}] 
 contract Victim {
     bool etherTransferred = false;
     //Attacker calls the withdraw() function to initiate the attack
     ~function withdraw()~{
     //Victim transfers ether which invokes the fallback function of the attacker
     if(etherTransferred ||
     `!msg.sender.call.value(1) ()`) throw;
     `etherTransferred = true;`
     }}
 contract Attacker{
    uint count = 0; 
    @function () payable{@
    if(++count < 10) Victim (msg.sender).withdraw();
    }}
\end{lstlisting}


\subsubsection*{Preventive Techniques}
Reentrancy vulnerability can be prevented by ensuring that state changing logic is committed before ether is sent out of the contract  through an external call. It is also a good coding practice to put any logic that performs external calls to unknown addresses at the last operation in a program's execution. This is known as the checks-effects-interactions pattern. Another technique is to use a mutex by adding a state variable which locks the contract during code execution, thus preventing re-entrant function calls.
\subsection{Out-of-Gas exception} 
The primitive function \textit{send} may cause an unexpected out-of-gas exception when transferring ether among contracts. 
There is a prefixed units of gas available to allow execution of a limited set of bytecode instructions and the call function will end up in an out-of-gas exception if not enough gas units are available.
\subsubsection*{Exploitation Case of Out-of-Gas Exception Vulnerability - King Of Ether Throne Attack}
The King of the Ether Throne contract ("KotET contract") is a game, where players compete to become the king by paying some ether as the claim price to the current king plus some fees to the contract owner \cite{surveyattacks}. After the contract declares a new King of the Ether Throne, the new claim price for the throne goes up by 50\%. When the KotET contract sent ether to the new King aspirant, it inadvertently included 2300 gas with the payment. As this was not enough gas to successfully process the payment and declare a new king, the wallet contract failed. This failure resulted in the ether being returned to the KotET contract. The KotET continued processing, thereby making the caller King despite the compensation payment not having been sent to the previous king (see listing \ref{kingofetherLst}).

\begin{lstlisting}[style=Solidity, basicstyle=\fontsize{7}{8}\selectfont, caption={Simplified Vulnerable King of Ether Attack - Out-Of-Gas Exception Vulnerability} ,label={kingofetherLst}] 
 contract KoEth {
     address public king;
     uint public claimPrize= 100;
     address owner;
     function KoEth () {
         owner = msg.sender;  king = msg.sender;}
     function () payable {
         ~if(msg.value < claimPrize) throw;~
        uint compensation = calculateCompensation();
        //"send" fails if the fallback function of reciever is expensive
        `king.send(compensation);`
        king = msg.sender; 
        claimPrize = calculateNewPrice(); }
        }
\end{lstlisting}

\subsubsection*{Preventive Technique}
This vulnerability can be prevented by using \textit{transfer()} function instead of \textit{send()} as the former will revert the local transactions if the external transaction reverts. However, if \textit{send()} is required then the return value of this function needs to be monitored. 
Another technique is to adopt a withdrawal pattern, wherein, each user is required to call an isolated function that manages ether transactions and does not affect the rest of the contract execution. Therefore, making the transaction management independent of the consequences of failed \textit{send()} transactions.
\subsection{Call to the unknown}
When a function invocation or an ether transfer unexpectedly invokes the fallback function of the callee/recipient. Some of the primitives of Solidity language that causes this are:
\begin{itemize}
    \item \textit{call} used to invoke a function or transfer ether
    \item \textit{send}, used to transfer ether from the running contract to some other contract
    \item \textit{delegatecall}, used to invoke a function or transfer ether in the caller environment
    \item direct call (see listing \ref{callToUnknownLst})
\end{itemize}

\begin{lstlisting}[style=Solidity,basicstyle=\fontsize{7}{8}\selectfont, caption={Call to the unknown - Direct Call},label={callToUnknownLst}]
 contract Alice{ function ping(uint) { returns (uint); }}
 contract Bob{ function pong (~Alice c~) { `c.ping (42);` }}
\end{lstlisting}

If an invoked function’s signature does not match with any existing function, then the call results in a call to recipient’s fallback function.
\subsubsection*{Exploitation Case of Call-to-Unknown Vulnerability - Second Parity MultiSig Wallet Attack}
On July 19, 2017, a major attack, in terms of Ether stolen, on the Ethereum network took place. The attacker’s account had drained 153,037 ETH from three high-profile multi-signature contracts used to store funds from past token sales \cite{ParityAttack}. The vulnerable MultiSig wallet was divided into two contracts(as shown in Listing \ref{MultiSigLst}) to reduce the size of each wallet and save gas:
\begin{itemize}
    \item A library contract called “WalletLibrary”,
    \item An actual “Wallet” contract
\end{itemize}
To begin with, the vulnerable contract had a simple constructor that delegates the initialization of the contract’s state to \textit{WalletLibrary}, followed by a  \textit{withdraw()} function which also delegates its execution to \textit{WalletLibrary}. Using these two functions, the attacker initiated two transactions to each of the vulnerable contracts: the first to obtain exclusive ownership of the MultiSig, and the second to transfer all of its funds to itself.
To obtain the ownership of the contract, the attacker needs to execute \textit{Wallet.initWallet(attacker)}. This triggers the \textit{Wallet}’s fallback function. In the \textit{Wallet}’s fallback function then initiates the \textit{delegatecall} in the \textit{WalletLibrary}. When \textit{WalletLibrary} receives the call, it finds that it's \textit{initWallet} function matches the function selector and runs \textit{initWallet(attacker)}. Thereby making the attacker the owner of the wallet and allowing him to be able to withdraw funds.
\begin{lstlisting}[style=Solidity,basicstyle=\fontsize{7}{8}\selectfont, caption={Simplified Vulnerable MultiSig Wallet \cite{ParityAttack}},label={MultiSigLst}] 
 contract WalletLibrary {
     address owner;
     ~function initWallet(address _owner) {~
         ~owner = _owner;~
     function changeOwner(address _new_owner) external {
         if (msg.sender == owner) {
             owner = _new_owner;}}
     function () payable { // receive money}
     function withdraw(uint amount) external returns (bool success) {
         @if (msg.sender == owner) {@
             `return owner.send(amount);`}
         else {
             return false;}}}
\end{lstlisting}
\subsubsection*{Preventive Technique}
Solidity has provision for implementing library contracts by using the keyword \textit{library} (see \cite{Solidity}). These library contracts are stateless and non-self-destructive. Forcing libraries to be stateless mitigates attacks whereby attackers modify the state of the library directly to affect the contracts that depend on the library's code. Therefore, when using \textit{call}, \textit{DelegateCall}, the call-to-the-unknown attack that may change the state of the victim contract can be prevented by building stateless libraries.
\subsection{Typecasts}
The fact that the Solidity compiler can detect some type errors may cause the programmers to believe that it also checks for the address of the contract being called and the interface declared by the caller function matches callee’s actual interface. The execution of a contract in the presence of such type mismatch errors will not throw exceptions at run-time and the caller is unaware of the error resulting in three different cases at run-time:
\begin{itemize}
    \item Incorrect contract address of callee function, the call returns without executing any code,
    \item Contact address of callee function matches with any other function’s signature, then that function is executed
    \item Contact address of callee function does not match with any function’s signature, then its fallback is executed.
\end{itemize}
\subsubsection*{Exploitation Case of Typecasts  Vulnerability - Reentrancy honey-Pot Attack}
Honey pot contracts are deployed on the Ethereum main network to capture Ethereum hackers who try to exploit the contracts. A small scale attack using the typecasts vulnerability was successfully launched on a honey pot developed to capture hackers trying to exploit the reentrancy vulnerabilities in smart contracts (see listing \ref{honeypotLst})
In listing \ref{honeypotLst}, this vulnerability can be exploited by replacing an expected contract address with a malicious address in the constructor.
\begin{lstlisting}[style=Solidity,basicstyle=\fontsize{7}{8}\selectfont, caption={Reentrancy Honey Pot Contract - Typecasts vulnerability},label={honeypotLst}] 
 contract Bank_Contract{
    mapping (address => uint) public balances;
    uint public MinDeposit = 1;
     ~function Bank_Contract(address _sender){~
     //update malicious address here
     }
    function sendDeposit() public payable{
        if(msg.value >= MinDeposit){
            balances[msg.sender] += msg.value;
            }}
    function withdraw(uint _am){
        if(_am <= balances[msg.sender]){
            @if(msg.sender.call.value(_am)())@{
                `balances[msg.sender] -= _am;`}}}
    function() public payable{}}
\end{lstlisting}
\subsubsection*{Preventive Technique}
 To prevent typecasting to malicious contract, the \textit{new} keyword can be used. This way an instance of the referenced contract cannot be changed without modifying the contract as this is created at deployment time. In listing \ref{newLst}, the constructor could be written like:
\begin{lstlisting}[style=Solidity,basicstyle=\fontsize{7}{8}\selectfont,  caption={Using \textit{new} to create an instance of a contract},label={newLst}] 
 constructor() {referenceContract = new reference(); }
\end{lstlisting}
Another technique is to hard code any external contract addresses in the contract to avoid malicious contracts getting referenced.
\subsection{Mishandled Exceptions}
There are many situations when an exception can be raised in Solidity but the way these exceptions are handled is not always the same. The exception handling is based on the interaction between contracts. This makes the contracts vulnerable to attacks because programmers will be unaware of any ether that is lost if these exceptions are not handled properly and the transactions are reverted. 
\begin{lstlisting}[style=Solidity,basicstyle=\fontsize{7}{8}\selectfont,  caption={function \textit{ping} of contract \textit{Alice} throws an exception},label={exception}] 
 contract Alice { 
 ~function ping(uint) {~
 // this function throws an exception 
 returns (uint);}}
 contract Bob { 
    uint x=0;
    function pong(Alice c){ x=1; `c.ping(42);` x=2;} }
\end{lstlisting}
In Listing \ref{exception}, the value of variable \textit{x} after the execution of contract \textit{Bob} varies depending on the method of the function call. If the \textit{ping} function of contract \textit{Alice} is called using a direct call, then the value of \textit{x} will be 0. Whereas, if the same function is called using the in-built function \textit{call} of Solidity, then  the value of \textit{x} will be 2. 
Moreover, in case of exceptions, if no bound is specified then all the available gas is lost.
\subsubsection*{Exploitation Case of Mishandled Exceptions Vulnerability - Governmental scheme  Attack}
(as shown in Listing \ref{governmentalLst})
In this attack, a contract that implements a flawed Ponzi scheme is targeted \cite{surveyattacks}. This attack is executed by exploiting the mishandled exceptions vulnerability in smart contracts. This scheme requires a participant to send a certain amount of ether to the scheme contract. If no one joins the scheme for 12 hours, the owner of the contract keeps his fee and transfers the remaining ether to the last participant. To join the scheme, a player must invest at least half of the claim prize. This claim prize increases upon each new investment. Anyone can invoke \textit{resetInvestment}, which transfers the claim prize (half of the invested total) to the last participant and sends the remaining ether to the contract owner. There is a key assumption in this contract that players are either users or contracts with empty fallback, and so will not cause an out-of-gas exception during send (as shown in Listing[\ref{governmentalLst}]

\begin{lstlisting}[style=Solidity,basicstyle=\fontsize{7}{8}\selectfont, caption={Simplified Governmental Attack - Mishandled Exceptions Vulnerability} ,label={governmentalLst}] 
 contract Governmental {
     address public owner;
     address public lastInvestor;
     uint public claimPrize= 1;
     function Governmental () {
         owner = msg.sender; 
        if(msg.value < 1) throw; }
     function invest (){
         ~if (msg.value < claimPrize/2) throw;~ 
        lastInvestor = msg.sender; 
        claimPrize += msg.value/2; }
     function () resetInvestment {
         `lastInvestor.send(claimPrize);` 
         //contract sends the prize money to the winner
         `owner.send(this.balance - 1);` 
         // and sends the remaining ether to the owner
         lastInvestor = 0; 
         claimPrize = 1;}
         }
\end{lstlisting}

\subsubsection*{Preventive Technique}
One technique to avoid this vulnerability would be to use one method of external call throughout. However, this is not an ideal preventive technique as different variations of an external call can be a necessity. Therefore, this vulnerability requires an update on the Solidity Language to make the consequences of a thrown exception uniform. 

\subsection{Weak Field Modifiers}
Fields in smart contracts can be labelled as \textit{Public} or \textit{Private}. However, these attributes are not enough to protect a field's value. This is because the default access modifier of afield in Solidity is \textit{public}. Whenever a field's value is changed, this change is published on the BT chain and there is a chance that an attacker would infer the changed value through previous hashes and new transaction hash. 
\begin{lstlisting}[style=Solidity,basicstyle=\fontsize{7}{8}\selectfont, caption={Simplified Multiplayer Games Attack - Weak Field Modifiers Vulnerability} ,label={multigameLst}] 
 contract OddsAndEvens {
    ~struct Player {address addr; uint number;}~
    ~Player[2] private players;~ 
    uint tot = 0; 
    address public owner;
    function OddsAndEvens () {
         owner = msg.sender; }
    function play (uint number){
        if (msg.value != 1 ) throw; 
        player[tot] = Player (msg.sender, number);
        tot++;
        if(tot == 2) winner(); }
    function winner() private{
        @uint n = players[0].number + players[1].number;@ 
        //contract sends 1.8 ether to the winner
        `players[n%2].addr.send(1.8);`
        delete players;
        tot = 0;}
    function getProfit(){
        //and sends remaining ether to the contract owner
         owner.send(this. balance);}
   }
\end{lstlisting}
\subsubsection*{Exploitation Case of Weak Field Modifier Vulnerability - Odd and Even Game Attack}
In this attack, a contract that implements a simple “odds and evens” game between two players is exploited \cite{surveyattacks}. An attacker impersonates the second player and when the first player makes his bet, the attacker infers this by BT network transactions. After inferring the first player's bet value, the attacker adjusts his bet accordingly that would guarantee his win. (as shown in Listing [\ref{multigameLst}])
\subsubsection*{Preventive Technique}
To avoid this smart contract's vulnerability, use the \textit{internal} modifier for functions instead of public. 
\subsection{Integer Underflow/Overflow Vulnerability}
An Integer overflow/underflow occurs when an arithmetic operation is performed that requires a fixed size variable to store data that falls outside the range of the variable’s data type. The EVM \cite{Ethereum} specifies data types with fixed-size for integers. Therefore, an integer variable can be represented by only a certain range of numbers. This vulnerability may be exploited by attackers by misusing the smart contract code and create unexpected logic flows.
\subsubsection*{Exploitation Case of Integer Underflow/Overflow Vulnerability - BECToken Attack}
On 22nd April 2018, there was an unusual token transfer in an \textit{ERC20 Smart contract} that prompted the contract owners to analyse the related smart contract code. The analysis resulted that the transfer was initiated as an “in-the-wild” attack that exploited the arithmetic overflow vulnerability in the contract. 
\subsubsection*{Preventive Technique}
This vulnerability can be avoided by Using  mathematical  libraries  instead  of the  standard  math  operations (addition, subtraction and multiplication). 
\subsection{DoS By An External Call Vulnerability}
When the flow of control is transferred to an external contract, the execution of the caller contract can fail accidentally or deliberately, which can cause a DoS state in the caller contract. The caller contract  can be in a DoS state when a transaction is reverted due to a failure in an external call, or the callee contract deliberately causes the transaction to be reverted to disrupt the execution of the caller contract.
\subsubsection*{Exploitation case of DoS By An External Call - HYIP (High Yield Investment Program)}
The contract \textit{HYIP} is yet another Ponzi scheme. This contract sends payments to lenders from funds collected via new lenders each day. The function \textit{sendPayment()} in Listing[\ref{HYIP}] contains the DoS by an external call vulnerability. The attack proceeds as follows: 
\begin{enumerate}
    \item The \textit{AttackerContract} lends funds to the \textit{HYIP} contract and throws an exception in its fallback function.
\item When function \textit{sendPayment()} is called to pay the
lenders, the fallback function of all the lenders is invoked and and the fall back function of this \textit{AttackerContract} throws an exception, causing a deliberate revert of the transaction and subsequently, a DoS to contract HYIP.
\end{enumerate}

\begin{lstlisting}[style=Solidity,basicstyle=\fontsize{7}{8}\selectfont,  caption={Contract HYIP - Exploited for DoS by an External Call Vulnerability},label={HYIP}] 
 contract HYIP {
    Lenders[] private lender;
    ~function sendPayment() {~
        for(uint i = lender.length; i−− > 0; ) {
            uint payment=(lenders[i].amount*/1000;
            `if(!lenders[i].addr.send(payment)) throw;`
            } }}
 contract AttackerContract {
    bool private attack = true;
    function() payable {
        `if (attack) throw;` 
        // callee fails the caller execution deliberately}
        }}
\end{lstlisting}
\subsubsection*{Preventive Technique}
This vulnerability exists because of inadequate exception handling around conditional and iteration statements. Placing any external calls initiated by a callee contract into a separate transaction can help reduce the damage caused by this vulnerability. Isolating statements with the following pattern can help avoid this vulnerability:
• an if-statement with an external function call in the condition and a throw or a revert in the body;
• a for- or an if-statement with an external function call in the condition.
Also, by asking the recipient to \textit{pull} funds out rather than sender using \textit{push} to send out funds. 
\begin{table*}
\begin{center}
\caption{Classification of Framework/Detection Tools Available for Vulnerabilities; {\color{red}{\xmark}}: None Available}
\label{ResearchDet}
\rotatebox[origin=c]{90}{ 
\subfloat[Classification of Framework/Detection Tools Available for Vulnerabilities]{
\scalebox{0.75}{
\begin{tabular}{|p{0.5cm}|p{2.5cm}|p{2.25cm}|p{2.25cm}|p{2.25cm}|p{2.25cm}|p{2.25cm}|p{2.25cm}|p{2.25cm}|p{2.25cm}|p{2cm}|p{2cm}|p{0.5cm}|} 
 \hline
\textbf{S.no.} &\textbf{Vulnerability} & \multicolumn{5}{c|}{\textbf{Static Analysis}} & \multicolumn{2}{c|}{\textbf{Dynamic Analysis}} & \multicolumn{3}{c|}{\textbf{Rectification provided at Level}}\\
\hline
&& \rotatebox[origin=c]{90}{\textbf{Symbolic. Execution}} & \rotatebox[origin=c]{90}{\textbf{CFG Construction}} & \rotatebox[origin=c]{90}{\textbf{Pattern Recognition}} & \rotatebox[origin=c]{90}{\textbf{Rule-Based Analysis}} &  \rotatebox[origin=c]{90}{\textbf{Decompilation Analysis}}& \rotatebox[origin=c]{90}{\textbf{Execution Trace at Runtime}} & \rotatebox[origin=c]{90}{\textbf{Fuzzing Transactions}} &
\rotatebox[origin=c]{90}{\textbf{Solidity Level}} &
\rotatebox[origin=c]{90}{\textbf{Bytecode Level}} & 
\rotatebox[origin=c]{90}{\textbf{Blockchain Level}}\\
\hline
1. & Re‐entrancy
& Oyente \cite{Oyente}, Mythril \cite{Mythril},  Securify \cite{Securify}, SmartCheck \cite{SmartCheck}
& Oyente \cite{Oyente}, ContractFuzzer \cite{ContractFuzzer},  SmartCheck \cite{SmartCheck}, SmartShield \cite{smartshield}
& Reguard \cite{Reguard},  Securify \cite{Securify},  SmartCheck \cite{SmartCheck}
& Mythril \cite{Mythril}, EthIR \cite{EthIR}
& Vandal \cite{Vandal}
& Reguard \cite{Reguard}, SmartShield \cite{smartshield}
& ContractFuzzer \cite{ContractFuzzer}, Ethploit \cite{Ethploit}
& SmartCheck \cite{SmartCheck}
& SmartShield \cite{smartshield}
& \color{red}{\xmark}\\
\hline

2. & Out-of-Gas - Failed Send Exception
& Mythril\cite{Mythril}
&\color{red}{\xmark} 
& \color{red}{\xmark}
& Mythril\cite{Mythril}
&  Vandal\cite{Vandal}
&\centering{\color{red}{\xmark}}
&\color{red}{\xmark}
&\color{red}{\xmark}
&\color{red}{\xmark}
&\color{red}{\xmark}\\
\hline

3. & Unpredictable state due to mishandled exceptions
& SmartCheck \cite{SmartCheck}, Securify \cite{Securify} 
& ContractFuzzer \cite{ContractFuzzer},  SmartCheck \cite{SmartCheck}
&  Securify \cite{Securify},  SmartCheck \cite{SmartCheck}
& EthIR\cite{EthIR}
& \color{red}{\xmark}
&\color{red}{\xmark}
& ContractFuzzer \cite{ContractFuzzer}
&\color{red}{\xmark}
&\color{red}{\xmark}
&\color{red}{\xmark}\\ 
\hline

4. & Call-to-Unknown 
& Mythril\cite{Mythril}, SmartCheck \cite{SmartCheck}, Securify \cite{Securify} 
&  SmartCheck \cite{SmartCheck}, SmartShield \cite{smartshield}
& Securify \cite{Securify} ,  SmartCheck \cite{SmartCheck}
& Mythril \cite{Mythril}
&  \color{red}{\xmark}
& MAIAN \cite{MAIAN}, SmartShield \cite{smartshield} 
& ContractFuzzer \cite{ContractFuzzer}, Ethploit \cite{Ethploit}  
& SmartCheck \cite{SmartCheck}
& SmartShield \cite{smartshield}
&\color{red}{\xmark}\\
\hline

5. & Typecast 
& \centering{\color{red}{\xmark}} 
&\color{red}{\xmark} 
& \color{red}{\xmark}
& \color{red}{\xmark}
& \color{red}{\xmark}
& \color{red}{\xmark}
& \color{red}{\xmark}
& \color{red}{\xmark}
&\color{red}{\xmark}
&\color{red}{\xmark}\\
\hline

6. & Weak Field Modifiers 
& SmartCheck \cite{SmartCheck}
&  SmartCheck \cite{SmartCheck}
&  SmartCheck \cite{SmartCheck}
& \color{red}{\xmark}
& \color{red}{\xmark}
& \color{red}{\xmark}
& Ethploit \cite{Ethploit}
& SmartCheck \cite{SmartCheck}
&\color{red}{\xmark}
&\color{red}{\xmark}\\
\hline

7. & Integer Underflow/OverFlow 
& Zeus \cite{Zeus}, Mythril \cite{Mythril}
&SmartShield \cite{smartshield}
& \color{red}{\xmark}
&  Mythril \cite{Mythril}
& \color{red}{\xmark}
& SmartShield \cite{smartshield}
& \color{red}{\xmark}
& SmartCheck \cite{SmartCheck}
& SmartShield \cite{smartshield}
&\color{red}{\xmark}\\
\hline

8. & DoS by an External Call 
&  SmartCheck \cite{SmartCheck}
& SmartCheck \cite{SmartCheck}, SmartShield \cite{smartshield}
&  SmartCheck \cite{SmartCheck}
& \color{red}{\xmark}
& \color{red}{\xmark}
& SmartShield \cite{smartshield}
& Ethploit \cite{Ethploit}
& SmartCheck \cite{SmartCheck}
& SmartShield \cite{smartshield}
&\color{red}{\xmark}\\
\hline
\end{tabular}
}}}
\qquad
\rotatebox[origin=c]{90}{ 
\subfloat[Usage of Ethereum Smart Contracts Format for Analysis by Frameworks/Detection Tools]{
\scalebox{0.75}{
\begin{tabular}{|p{0.5cm}|p{3.25cm}|p{3cm}|p{3cm}|p{3cm}|p{3cm}|} 
\hline 
\textbf{S.no.} &\textbf{Framework/Detection Tool} & \textbf{Abstract Syntax Tree} & \textbf{Solidity Code} & \textbf{Intermediate Representation of Solidity code} & \textbf{Bytecode}\\
 \hline
 1. & Mythril & \color{red}{\xmark} & \color{red}{\xmark} &\color{red}{\xmark} & \color{blue}{\cmark} \\ 
 2. & Zeus & \color{blue}{\cmark} & \color{red}{\xmark} &\color{red}{\xmark} & \color{red}{\xmark}\\
 3. & Oyente & \color{blue}{\cmark} & \color{red}{\xmark} &\color{red}{\xmark} & \color{blue}{\cmark}  \\
 4. & ContractFuzzer & \color{blue}{\cmark} & \color{red}{\xmark} &\color{red}{\xmark} & \color{blue}{\cmark} \\
 5. & EthIR  & \color{blue}{\cmark} & \color{red}{\xmark} &\color{red}{\xmark} & \color{blue}{\cmark}  \\
 6. & MAIAN & \color{blue}{\cmark} & \color{red}{\xmark}& \color{red}{\xmark} & \color{red}{\xmark} \\
 7. & SmartCheck & \color{blue}{\cmark} & \color{blue}{\cmark} &  \color{red}{\xmark} & \color{blue}{\cmark}  \\
 8. & Reguard & \color{blue}{\cmark} & \color{red}{\xmark} & \color{blue}{\cmark} & \color{red}{\xmark}  \\
 9. & Securify & \color{red}{\xmark} & \color{blue}{\cmark} & \color{blue}{\cmark} & \color{blue}{\cmark} \\
 10. & Vandal& \color{blue}{\cmark} & \color{red}{\xmark} &\color{red}{\xmark} & \color{blue}{\cmark}  \\
 11. & SmartShield & \color{red}{\xmark} & \color{red}{\xmark} & \color{blue}{\cmark} & \color{blue}{\cmark}  \\
 12. & Ethploit& \color{red}{\xmark} & \color{blue}{\cmark} & \color{red}{\xmark} & \color{red}{\xmark}  \\
 \hline
\end{tabular}
}}}
\qquad
\end{center}
\end{table*}

\section{Research Analysis and Insights}
There has been extensive research going on to identify, characterize and prevent vulnerabilities in Ethereum Smart Contracts. For this paper, we considered the following research studies,
\begin{itemize}
    \item Luu et al.\cite{Oyente} developed the Oyente analyzer that performs symbolic execution on contract functions and identifies vulnerabilities based on simple patterns. According to this framework, the vulnerabilities are classified into the following groups: transaction-ordering dependent, timestamp dependence, re-entrance handling, and mishandled exceptions.
    \item SmartCheck \cite{SmartCheck} is a pattern-based analysis tool that uses XPath to detect if any vulnerabilities pattern exists in a Smart Contract. To do so, it transforms the Smart Contract into XML representation.
    \item ReGuard \cite{Reguard} is a combined static and dynamic analysis tool to detect reentrancy vulnerabilities in Smart-Contracts developed by Liu et al.\cite{Reguard}. This tool tests the Smart-Contracts by initially transforming the Smart-Contract code into C++ and then generating fuzzing inputs to recreate Blockchain transactions as possible attacks. Then, ReGuard performs vulnerability detection through dynamic analysis.
    \item Contract Fuzzer \cite{ContractFuzzer} is a tool developed by Jiang et al. that tests the Smart-Contracts for identifying vulnerabilities in them by using the fuzzing technique. To detect the vulnerabilities this tool starts with an initial analysis of the interfaces that the Smart-Contract exposes, it then randomly develops fuzzing inputs for these interfaces and observes the execution logs of the application.
    \item Mythril \cite{Mythril} is a command-line tool in Python developed by ConsenSys for analyzing smart contracts interactively. It executes EVM bytecode symbolically and represents it in the form of a CFG, with the nodes containing disassembled code and the edges being labelled by path formulas.
    \item MAIAN \cite{MAIAN} is a python based tool that uses Oyente \cite{Oyente} for the detection of vulnerabilities that require multiple transactions. It executes EVM bytecode symbolically and checks for execution traces. To discard false positives, the contracts are dynamically analyzed by deploying them on a private blockchain and attacking them with the computed transactions.
    \item Securify \cite{Securify} uses EVM bytecode and security properties of a smart contract as inputs. A Security property consists of compliance and violation patterns. This tool uses the decompilation analysis method  and represents the code as Data Log facts. This framework infers that if a pattern is detected, then the code possesses the corresponding security vulnerability.
    \item Vandal \cite{Vandal} is a command-line tool written in Python which disassembles and decompiles EVM bytecode into an intermediate representation and constructs a CFG.
    \item Zeus \cite{Zeus} is a tool developed by IBM Research India. Similar to Securify \cite{Securify}, this tool takes Solidity code and policies as input. These policies are checks that specify if the code meets a safety property expressed in the policy. Zeus converts Solidity code into LLVM bitcode, which is then instrumented with assertions corresponding to the policy.
    \item EthIR \cite{EthIR} is written in Python and analyzes only particular versions of the Solidity compiler, and Go-Ethereum. This framework transforms bytecode into an intermediate representation compatible with a static analyzer built by the same developers as of EthIR. This framework extends Oyente \cite{Oyente}. The CFG is represented as guarded rules and this rule-based representation is then supplied as an input to the general purpose static analyzer.
\end{itemize}

\begin{figure*}
    \centering{\includegraphics[width=1\linewidth]{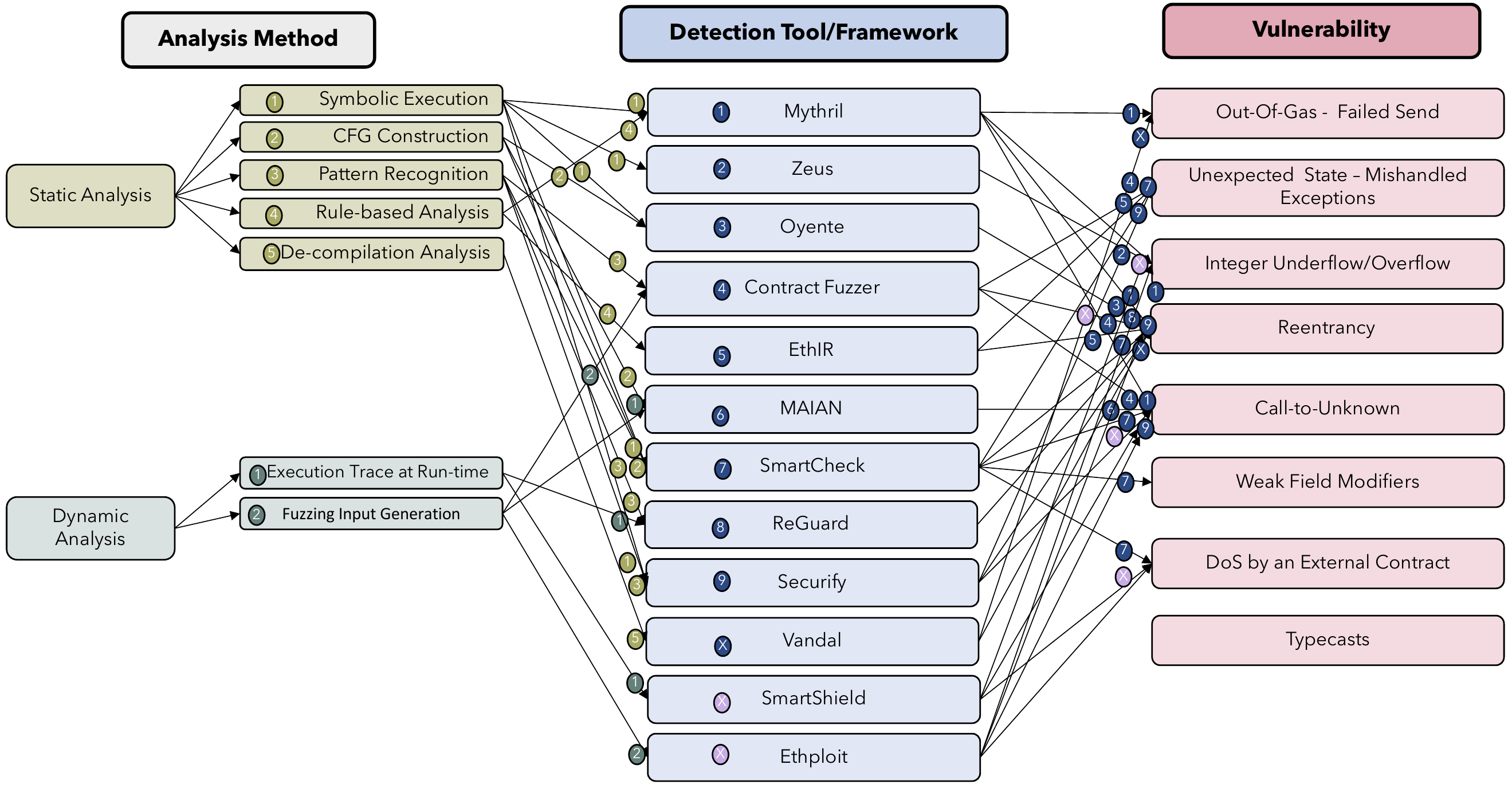}}
    \caption{Relationship between Framework/Detection Tools Available and Vulnerabilities}
    \label{fig:relationship}
\end{figure*}

\begin{figure*}
    \centering
     \subfloat[Most Severe Attack in terms of Ether lost]{
     {\includegraphics[width=9.5cm]{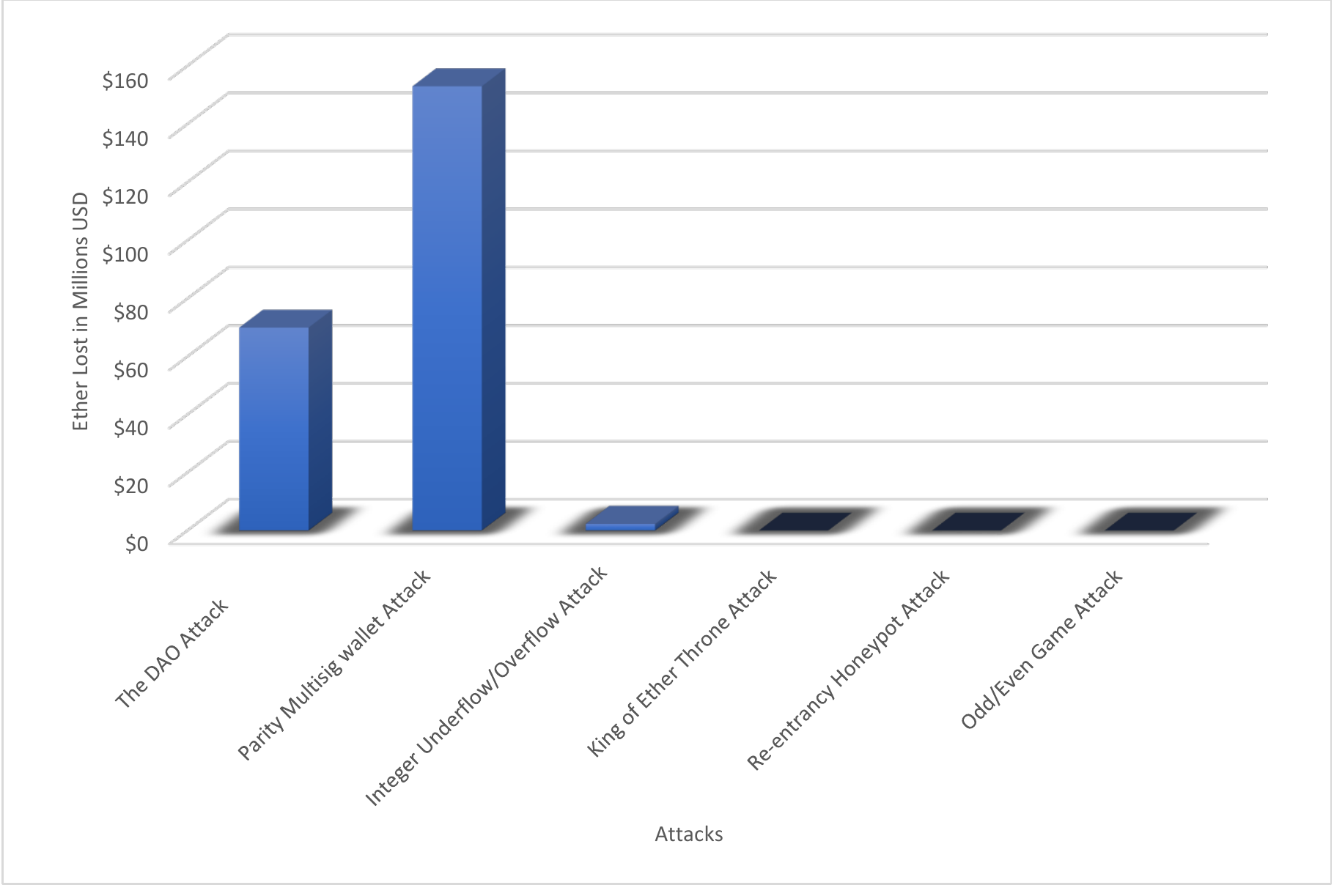} }}
     \subfloat[Most Researched Vulnerability]{
     {\includegraphics[width=9.5cm]{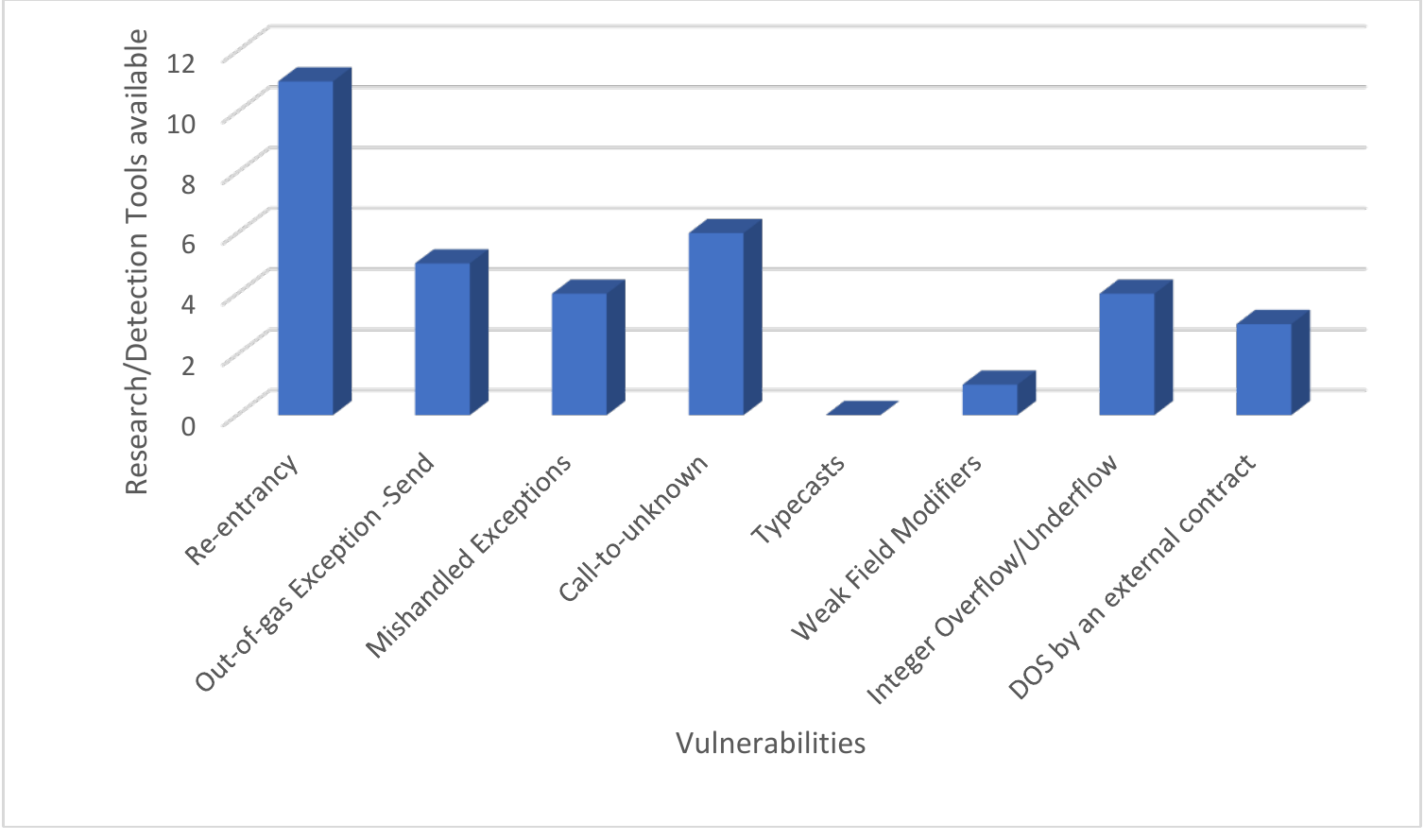}}}
    \caption{Research Statistics of vulnerabilities}
    \label{fig:researchStatistics}
\end{figure*}

Figure[\ref{fig:researchStatistics}] shows the statistics surrounding the research of individual vulnerabilities. It is evident from Figure[\ref{fig:researchStatistics}] that the reentrancy vulnerability has been talked about the most and there has been almost no research related to typecasts vulnerability. However, Figure[\ref{fig:researchStatistics}] suggests that the most amount of Ether was lost due to Parity Multisig Wallet Attack which was an exploitation of this vulnerability that cost ~\$150 Million worth of Ether, followed by the DAO attack which was an exploitation of reentrancy vulnerability that cost ~\$70 Million worth of Ether.
Literature review of the selected detection tools/frameworks for this paper shows that these tools/frameworks adopted one of the analysis methodologies mentioned in the background section of this paper to analyse Ethereum Smart Contracts
Figure [\ref{fig:relationship}] illustrates the adopted analysis methodology by various research frameworks and detection tools for each vulnerability respectively.

The reentrancy vulnerability's statistics illustrated in Figure[\ref{fig:relationship}] show that most of the frameworks and detection tools Oyente \cite{Oyente}, Mythril \cite{Mythril}, SmartCheck \cite{SmartCheck}, Vandal \cite{Vandal}, Securify \cite{Securify}, EthIR \cite{EthIR} surveyed in this paper, adopted the static analysis method to detect this vulnerability in smart contracts. Static analysis methods can detect the existence of the pattern defined for this vulnerability, however, defining the pattern of this vulnerability is also a challenge. The confirmation of the existence of this vulnerability can be more accurately outlined by a successful reentrancy generating transaction from an external contract to the contract under test. Only two of the analyzed research works, ContractFuzzer \cite{ContractFuzzer}, Reguard \cite{Reguard}, utilized the combined static and dynamic analysis method which is believed to be a better analysis methodology for this vulnerability. 

The out-of-gas due to failed send vulnerability's statistics depicted in Figure[\ref{fig:relationship}] shows that only static analysis methods were adopted by detection tools/frameworks to detect this vulnerability \cite{Vandal}, \cite{Mythril}. 

Figure[\ref{fig:relationship}] shows that the vulnerability caused by mishandled exceptions where the state of a smart contract becomes unpredictable was found to be identified mostly by using static analysis SmartCheck \cite{SmartCheck}, EthIR \cite{EthIR}, Securify \cite{Securify}. However, ContractFuzzer \cite{ContractFuzzer} also successfully detects this vulnerability using the fuzzing technique to generate multiple transaction scenarios. 
The Call-to-Unknown vulnerability was found to be detected by using combined static and dynamic analysis approach by two of the detection tools surveyed \cite{ContractFuzzer}, \cite{MAIAN}. (See Figure[\ref{fig:relationship}])
The weak field modifiers vulnerability was addressed by only one vulnerability detection tool \cite{SmartCheck} (See Figure[\ref{fig:relationship}]. The vulnerability caused due to unchecked math or more specifically Integer underflow/overflow ignorance was detected by two of the detection tools \cite{Zeus}, \cite{Mythril}. Whereas, none of the detection tools had an analysis method to detect the vulnerability caused due to typecasting in Solidity. 
\section{Conclusion}
This paper presents an analysis of the security vulnerabilities of Ethereum smart contracts, real-world exploitation cases of these vulnerabilities and their preventive techniques. Our paper targets eight security vulnerabilities in Blockchain 2.0 applications, specifically in Ethereum Smart Contracts. The vulnerabilities discussed are at the level of the application layer. The preventive techniques thus require alterations at the programming level. The research analysis and insights provided in this paper aim at directing the future study in this field towards the development of more robust vulnerabilities detection tools. Our analysis is based on 
\begin{itemize}
    \item the growing academic literature on the topic,
    \item the discussion forums and Internet blogs of smart contracts programmers.
\end{itemize}
\section*{Acknowledgments}
This work is supported in part by the Natural Sciences and Engineering Research Council of Canada (NSERC).
\nocite*
\bibliographystyle{ACM-Reference-Format}
\bibliography{Noama-Cascon}
\end{document}